\documentclass[aip,jcp,reprint,amsmath,amssymb,groupedaddress,superscriptaddress]{revtex4-1}
\usepackage{color}
\usepackage{enumitem}
\usepackage{graphicx}
\usepackage[normalem]{ulem}
\usepackage{comment}

\newif\ifHighlitedChanges
\def\ifHighlitedChanges{\iftrue}
\ifHighlitedChanges
  
  \def\STRIKE#1{{\color{red}\sout{#1}}}
\else
  
  \def\STRIKE#1{\relax}
\fi

\begin{document}

\bibliographystyle{apsrev}

\title{Electron hopping heat transport in molecules}
\author{Galen T. Craven}
\affiliation{Theoretical Division, Los Alamos National Laboratory, Los Alamos, New Mexico 87544} 
\author{Abraham Nitzan}
\affiliation{Department of Chemistry, University of Pennsylvania, Philadelphia, PA  19104, USA} 
\affiliation{School of Chemistry, Tel Aviv University, Tel Aviv 69978, Israel}

\begin{abstract}
The realization of single-molecule thermal conductance measurements has driven the need for theoretical tools to describe conduction processes that occur over atomistic length scales.
In macroscale systems, the principle that is typically used to understand thermal conductivity is Fourier’s law.
At molecular length scales, however, deviations from Fourier's law are common in part because microscale thermal transport properties typically depend on the complex interplay between multiple heat conduction mechanisms.
Here, the thermal transport properties that arise from electron transfer across a thermal gradient in a molecular conduction junction are examined theoretically.
We illustrate how transport in a model junction is affected by varying the electronic structure and length of the molecular bridge in the junction as well as the strength of the coupling between the bridge and its surrounding environment.
Three findings are of note: 
First, the transport properties can vary significantly depending on the characteristics of the molecular bridge and its environment;
second, the system's thermal conductance commonly deviates from Fourier's law;
and third, in properly engineered systems, the magnitude of electron hopping thermal conductance is similar to what has been measured in single-molecule devices.
\end{abstract}
   \maketitle

\section{Introduction}	
The development of experimental set-ups capable of single-molecule thermal conductance measurements has opened the possibility for advanced understanding of the interplay between electronic and heat transport at the molecular level. \cite{Reddy2019nature,Mosso2019,Reddy2007,Wang2020review,Gehring2021}
The thermal conductance of molecular devices is typically investigated in order to optimize operational stability of nano-electronics or to control the functionality in thermal and thermoelectric devices. \cite{Reddy2007,Wang2020review,Gehring2021,Malen2009,Ke2009,Tan2011,Dubi2011,Li2012,Maldovan2013,Luo2013,Lee2013,Kim2014,Lee2014,Leitner2015,Segal2016,Leitner2016,Reddy2019nature,Mosso2019}
One of the major problems in examining thermal conductance at the nanoscale is that thermal fluctuations can significantly affect the stability and functionality of small systems, making single-molecule experimental measurements difficult to perform.
A system that is broadly used in the analysis of nanoscale thermal and electrical conductivity is a molecular junction---a device consisting of a molecular bridge seated between two electrodes. \cite{Ratner1974rectifier,Nitzan2003electron,Ratner2013review}
Molecular junctions are central in the design of many nanoscale technologies, particularly molecular electronics, and have therefore been studied extensively both theoretically and experimentally. \cite{Reddy2007,Tan2011,Lee2013,Kim2014,Venkataraman2015,Garner2018,Reddy2019nature,Mosso2019}
Applying either a temperature bias or a voltage bias, or both, across a molecular junction induces an electric and/or heat current through the molecular bridge which enables the control of conductivity at the molecular level, 
a significant step toward the realization of technological devices at the smallest level of miniaturization.

At the macroscale, a fundamental result that is broadly applied to analyze heat transport is Fourier's law:
\begin{equation}
\label{eq:FL}
\mathbf{J}_\mathcal{Q}= -\kappa \nabla T,
\end{equation}
which states that the heat current density $\mathbf{J}_\mathcal{Q}$ generated in a system in response to a temperature gradient $\nabla T$ is proportional to the system's thermal conductivity $\kappa$. \cite{Dubi2011,Segal2016} 
Fourier's law is applied in the heat transport analysis of myriad and diverse macroscopic systems. 
Despite its importance, however, the question of how Fourier's law arises from atomistic interactions at the microscale is currently unresolved.\cite{BonettoFourier2000,Bonetto2004SCR,Segal2009SCR,Chang2008} 
Fourier's law is a limiting behavior of atomistic dynamics and the physics underlying the way this limit is reached under different environmental conditions is not universal.
While progress has been made toward a full understanding of the molecular origins of Fourier's law, 
the methods and models used in these studies typically rely on imposed environmental conditions that are designed specifically to generate Fourier's law behavior,
and thus the fundamental question of how Eq.~(\ref{eq:FL}) manifests at the microscale remains open.

In purely molecular systems, phononic (vibrational) heat transport is traditionally thought to be the sole thermal conductance mechanism.\cite{Velizhanin2015,Dhar2015,Li2012,Li2006,Chang2006,He2016,Leitner2008,Leitner2013,Leitner2016} 
Recently, however, another heat transport mechanism, termed electron-transfer-induced heat transport (ETIHT) has been identified in molecular systems.\cite{craven16c, craven21a}
This heat transport is generated by the transfer of electrons across a thermal gradient between molecules.
Theoretical studies have suggested that the magnitude of ETIHT can be similar to phononic heat current and thus contribute significantly to the overall thermal conductance in molecular systems.  \cite{craven17e}

In this article, we examine the electron-transfer-induced (ETI), i.e., electron hopping, thermal conductance properties of a model molecular junction consisting of a molecular bridge of multiple charge transfer sites seated between two metal electrodes, 
where each site in the bridge is characterized by a different local temperature.
This article advances our previous work on electron hopping thermal and thermoelectric transport \cite{craven16c,matyushov16c,craven17a,craven17b,craven17e,craven18b,craven20a} by examining how the magnitude of ETIHT depends on the length of the molecular bridge in a molecular junction and by performing a detailed analysis of how variations in electronic properties and electron-phonon interactions affect ETIHT.
The dominant transport channels in molecular junctions often have quasi one-dimensional topology \cite{Nitzan2003thermal} and therefore, in this model, thermal conductance $\mathcal{K}$ replaces thermal conductivity as the pertinent transport property.
Under linear response conditions, the total thermal conductance can be defined as
\begin{equation}
\label{eq:themalcond}
\mathcal{K} = \!\lim_{\Delta T \to 0}\frac{\mathcal{J}_\mathcal{Q}}{\Delta T} \bigg|_{\mathcal{J}_\text{el} = 0},
\end{equation}
where $\mathcal{J}_\mathcal{Q}$ is the total heat current through the junction calculated under zero electronic current conditions $\mathcal{J}_\text{el} = 0$. 

In molecular junctions, a simple picture of the total thermal conductance $\mathcal{K}$ can be constructed by taking a sum over four major contributions: 
$\mathcal{K} =\mathcal{K}_\text{el} + \mathcal{K}_\text{rad}+ \mathcal{K}_\text{ph} + \mathcal{K}_\text{ETI}$, 
where $\mathcal{K}_\text{el}$ is the electronic thermal conductance, $\mathcal{K}_\text{rad}$ is the thermal conductance due to radiative heat transfer between electrodes, and $\mathcal{K}_\text{ph}$ and $\mathcal{K}_\text{ETI}$ are, respectively, the thermal conductances due to phononic and ETI heat transport.
What contribution each mechanisms makes to the total thermal conductance will varying depending on the properties of the molecular bridge, its surrounding environment, and the composition and geometry of the electrodes.
In this article, we disregard the phononic and radiative contributions in our calculations in order to concentrate on the other mechanisms.
Although it is formally disregarded in our calculations, comparisons will be made throughout between literature values for the phononic conductance and the thermal conductances calculated here. 
The ETI contribution to the total thermal conductance is examined by applying previously developed theoretical formalisms.\cite{craven16c,matyushov16c,craven17a,craven17b,craven17e,craven18b,craven20a}

The remainder of this article is organized as follows:
Section~\ref{sec:model} contains details of both the model and the formalism that are applied to examine ETI thermal transport in molecular junctions.
Specifically, expressions are given for the electric current and ETI heat current in a model molecular conduction junction.
These expressions are used to evaluate the system's thermal properties under different conditions. 
The numerical procedures used to evaluate these currents are also discussed. 
Section~\ref{sec:ballistic} contains the theoretical formalism used to examine the same transport properties generated by ballistic electron transport.
In Sec.~\ref{sec:results}, the results of this model are presented, focusing on analyzing the thermal response trends as function of temperature, electronic structure of the bridge, and molecular bridge length.
The thermal conductance values obtained in the hopping inelastic ETI heat transport limit are compared with those obtained in the ballistic elastic transport limit.
Conclusions and future directions are discussed in Sec.~\ref{sec:conc}.

\section{Model details \label{sec:model}}

We consider a physical model used previously by us in Refs.~\citenum{craven17a}  and \citenum{craven20a} that consists of a molecular bridge seated between two metal electrodes. For completeness of exposition in this work, we reexamine the details of this model below.

\begin{figure}[t]
\includegraphics[width = 8.5cm]{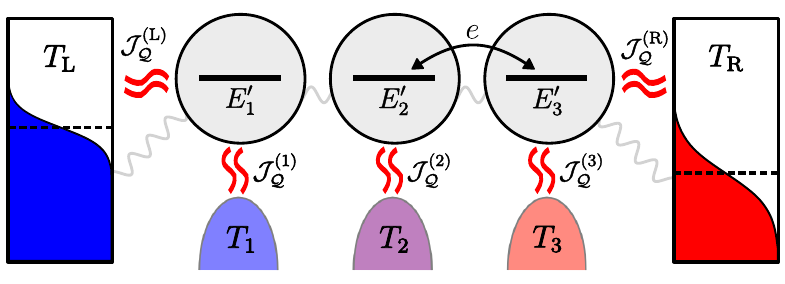}
\caption{\label{fig:Main_final}
Diagram of a representative molecular junction comprising three charge transfer sites (gray circles) seated between two metal electrodes (rectangles).
Each molecular site $s$ has an electronic occupation energy $E'_s$ and is in contact with a local thermal reservoir (semi-ellipse) with temperature $T_s$.
The wavy gray lines represent molecule-molecule and molecule-metal bonding.
The wavy red lines denote heat currents.
The electrodes have respective temperatures $T_\text{L}$ and $T_\text{R}$.
In the diagram of each electrode, the dashed line denotes the Fermi level and the colored region represents the corresponding Fermi-Dirac distribution.
A heat current $\mathcal{J}^{(s)}_\mathcal{Q}$ is associated with each thermal reservoir. 
The heat currents associated with the molecule-metal interfaces are $\mathcal{J}^{(\text{L})}_\mathcal{Q}$ and $\mathcal{J}^{(\text{R})}_\mathcal{Q}$.
}
\end{figure}

The molecular bridge is comprised of $N$ charge transfer sites, labeled $1,2, \ldots, N$.
The electrodes are are denoted as $\text{M}_\text{L}$ (left electrode) and $\text{M}_\text{R}$ (right electrode).
Each molecular charge transfer site $s$ in the bridge is associated with an electronic occupation energy $E_s'$, a local thermal environment at temperature $T_s$, and a set of vibrational modes that are coupled to the ET process. 
These modes are assumed to be in equilibrium at site's local temperature. \cite{note1}
For simplicity, herein, we consider the case of a single vibrational mode $x_s$ associated with each site $s$,
although it is straightforward to generalize this model to the case of multiple modes. \cite{craven17b}
This 
assumption
does not affect the general observations about thermal conductance that will be made later.
Each vibrational mode can be thought of as a collective variable that parameterizes all of the vibrational motions associated with the environment of its respective site---a picture that is similar to the typical Marcus picture of ET.
The left and right electrodes are assumed to be in both thermal and electrochemical equilibrium.
These equilibria are characterized by respective temperatures $T_\text{L} = T +\Delta T/2$ and $T_\text{R} = T -\Delta T/2$ and chemical potentials $\mu_\text{L} = \mu - e V/2$ and $\mu_\text{R} = \mu + e V/2$,
where $T$ is temperature and $\mu$ is the Fermi level. 
We disregard any temperature dependence of the Fermi levels.
The potential bias across the device is $\Phi = e V$.
The spatial distribution of the energy levels $E'_s$ in the junction, i.e., the electronic structure of the molecular bridge, is termed an energy landscape $\mathcal{L}$.
Varying $\mathcal{L}$ by modifying the electronic properties of the bridge alters the conduction properties of the device.
A schematic diagram of the molecular junction model is shown in Fig.~\ref{fig:Main_final}.

In the hopping limit of electron transport, 
the vibrational environment associated with the molecular bridge is strongly coupled with the electronic motion through the junction.
In this limit, electronic conduction proceeds through an inelastic mechanism in which the charge density of an electron moving through the junction localizes on a particular site in the bridge.
The system can therefore be described by $N+1$ electronic states $a \in \left\{\text{M},1, \ldots, N\right\}$ where each state $a \neq \text{M}$ corresponds to electron density being localized on molecular site $a$.
The state in which the electron occupies an energy level in one of the metal electrodes is denoted by $a  = \text{M}$ .
We consider only singly-occupied sites, the rationale being that a Coulomb blockade makes sites with multiple electron occupancy too high in energy.
For this reason, multiple-occupancy states are usually disregarded when modeling molecular conduction junctions, and we also disregard them here.

Electron hopping dominated by electron-phonon interaction that arises from electron localization on each site can be described using the Marcus formalism \cite{craven16c, Marcus1956,Hush1961,Kuznetsov1999Electron,Nitzan2006chemical}
in which the energy of each state $a$ is described using a paraboloid free energy surface: 
\begin{equation}
\label{eq:energysurf}
\begin{aligned}
&E_a\left(x_1,\ldots,x_N\right) = \sum\limits^N_{{s=1}} \frac{1}{2}k_s\left(x_s - \bar{x}^{(a)}_s\right)^2  + E'_a,
\end{aligned}
\end{equation}
where $k_s$ is the force constant of mode $x_s$,
$\bar{x}^{(a)}_s$ is a shift in mode $x_s$ in response to electron localization on site $a$,
and $E_a^{'}$ is the electronic energy origin of state $a$.
The main approximation involved in using the Marcus formalism to describe sequential electron hopping is the assumption that thermal equilibration is fast relative to the time between subsequent hopping events.
A generalization to multithermal situations when this approximation is not invoked was recently been described. \cite{Henning2020,Henning2022}
We apply the displaced harmonic oscillator approximation and therefore take the force constant to be independent of the electronic state of the system.
In the particular case of state $a = \text{M}$, the energy origin is $E'_\text{M} = \mu$.
The reorganization energy of mode $s$ over a specific state transition $a \to b$ is
\begin{equation}
\lambda^{(a,b)}_{s} =  \frac{1}{2} k_s \left(\bar{x}_s^{(a)}-\bar{x}_s^{(b)}\right)^2,
\end{equation}
and the total reorganization energy of the entire system over the same transition is
\begin{equation}
\lambda^{(a,b)} = \sum_{s=1}^N \lambda^{(a,b)}_{s}.
\end{equation}
In all the numerical results we present (See Sec.~\ref{sec:results}), the total reorganization energy and the reorganization of every site are taken to be the same for each state transition. 
The total reorganization energy can therefore be defined simply as $\lambda$.

\subsection{Molecule-to-molecule electron transfer rates}

The ET rate across the thermal gradient between sites in the molecular bridge is \cite{craven16c}
\begin{align}
\label{eq:genrate}
\nonumber k_{a \to b}  &= \frac{|V_{a,b}|^2}{\hbar}\sqrt{\frac{\pi}{k_\text{B}\left(T_a \lambda^{(a,b)}_{a} + T_b \lambda^{(a,b)}_{b}\right)}} \\[0ex]
& \qquad \times \exp{\left[ -\frac{\left(\Delta E_{ba} + \lambda^{(a,b)}\right)^2}{4 k_\text{B} \left(T_a \lambda^{(a,b)}_{a} + T_b \lambda^{(a,b)}_{b} \right)} \right]},
\end{align}
where $V_{a,b}$ is the electronic coupling between states $a$ and $b$ and $\Delta E_{ba} = -\Delta E_{ab}  = E'_b-E'_a$ is the reaction free energy.
Equation~(\ref{eq:genrate}) is a bithermal reaction rate, previously derived by us,\cite{craven16c,craven21a} that describes ET in the presence of a temperature difference between molecules.
In the unithermal limit where $T_a = T_b = T$, it reduces to the traditional Marcus ET rate expression that describes phonon-assisted electron hopping between molecular sites in equilibrium in the limit of strong electron-phonon coupling and high temperature.

\subsection{Molecule-to-metal electron transfer rates}	

In the high-temperature limit, the ET rates across the temperature gradients between molecule and metal at the left and right electrodes are, \cite{craven17a}
\begin{align}
\label{eq:moltometalL}
 &\nonumber k_{1 \to \text{M}_\text{L}} = \sqrt{\frac{1}{4 \pi k_\text{B} T_1 \lambda^{(1,\text{M})}_{1}}} \int_{-\infty}^{\infty}  \big[1-f\left(T_\text{L}, \mu_\text{L}, \epsilon\right)\big]  \\
& \qquad \times \Gamma_\text{L}(\epsilon)  \exp\Bigg[- \frac{\big(\Delta E_\text{L} +\epsilon+\lambda^{(1,\text{M})}_{1}\big)^2}{4 k_\text{B} T_1 \lambda^{(1,\text{M})}_{1}}\Bigg]\,d\epsilon, \\[2ex]
\label{eq:moltometalR}
& \nonumber k_{N \to \text{M}_\text{R}} = \sqrt{\frac{1}{4 \pi k_\text{B} T_N \lambda^{(N,\text{M})}_{N}}} \int_{-\infty}^{\infty}  
\big[1-f\left(T_\text{R}, \mu_\text{R}, \epsilon\right)\big]  \\
& \qquad \times \Gamma_\text{R}(\epsilon)  \exp\left[- \frac{(\Delta E_\text{R} +\epsilon+\lambda^{(N,\text{M})}_{N})^2}{4 k_\text{B} T_N  \lambda^{(N,\text{M})}_{N}}\right]\,d\epsilon,
\end{align}
and the corresponding rates of ET from metal to molecule are
\begin{align}
\label{eq:metaltomolL}
& \nonumber k_{\text{M}_\text{L} \to 1} = \sqrt{\frac{1}{4 \pi k_\text{B} T_1 \lambda^{(\text{M},1)}_{1}}} \int_{-\infty}^{\infty}  f\left(T_\text{L}, \mu_\text{L}, \epsilon\right)  \\
& \qquad \times \Gamma_\text{L}(\epsilon)  \exp\left[- \frac{(-\Delta E_\text{L} -\epsilon+\lambda^{(\text{M},1)}_{1})^2}{4 k_\text{B} T_1 \lambda^{(\text{M},1)}_{1}}\right]\,d\epsilon, \\[2ex]
\label{eq:metaltomolR}
& \nonumber k_{\text{M}_\text{R} \to N} = \sqrt{\frac{1}{4 \pi k_\text{B} T_N \lambda^{(\text{M},N)}_{N} }} \int_{-\infty}^{\infty}  
f\left(T_\text{R}, \mu_\text{R}, \epsilon\right)   \\
& \qquad \times \Gamma_\text{R}(\epsilon) \exp\left[-\frac{(-\Delta E_\text{R} -\epsilon+\lambda^{(\text{M},N)}_{N})^2}{4 k_\text{B} T_N \lambda^{(\text{M},N)}_{N}}\right]\,d\epsilon,
\end{align}
where $f(T,\mu,\epsilon) = (\exp\left[(\epsilon-\mu) / k_\text{B} T\right]+1)^{-1}$ is the Fermi-Dirac distribution. 
The molecule-metal coupling strength at electrode \text{E} is defined by
\begin{equation}
\Gamma_\text{E}(\epsilon) = \bigg(\frac{2 \pi}{ \hbar} |V_\text{E}|^2 \rho_\text{E}\bigg)_{\!\epsilon} : \text{E} \in \left\{\text{L}, \text{R}\right\}.
\end{equation}	
These rates can be derived from a transition state theory approach\cite{Nitzan2006chemical,craven15c} using assumptions for the transmission factor \cite{craven17a} or from a Fermi's golden rule formulation. \cite{craven21a}
Note that the single electron density of states $\rho_\text{E}$ and electronic tunneling coupling $V_\text{E}$ between molecule and metal both formally depend on the energy level $\epsilon$, 
however, in all numerical calculations we assume them to be independent of energy using the so-called wide-band approximation.
The reaction free energies for state transitions $1 \to \text{M}$ at the left electrode and $N \to \text{M}$ at the right electrode are
\begin{equation} 
\Delta E_\text{L} =  E'_{\text{M}} - E'_1,
\end{equation}
and 
\begin{equation} 
\Delta E_\text{R} = E'_{\text{M}} - E'_N,
\end{equation}
respectively, 
where $E'_\text{M} = \mu$.
In the limit in which the temperature bias vanishes, the rate expressions in Eqs.~(\ref{eq:moltometalL}) - (\ref{eq:metaltomolR}) are the Marcus-Hush-Chidsey rates for ET between a molecular and metal under equilibrium conditions. \cite{Marcus1965,Hush1968,Chidsey1991,Nitzan2006chemical,Nitzan2011,Compton2012}

\subsection{Electric current}

An expression for the electric current $\mathcal{J}_\text{el}$ through the junction can be derived from the kinetic master equations describing the probability $\mathcal{P}_a$  that the system is in each state $a \in \left\{\text{M},1, \ldots, N\right\}$:
\begin{equation}
\begin{aligned}
\label{eq:occ2}
\dot{\mathcal{P}}_{\text{M}} &= - k_{\text{M}_\text{L} \to 1} \mathcal{P}_{\text{M}}+ k_{1 \to \text{M}_\text{L}} \mathcal{P}_1\\
&\qquad - k_{\text{M}_\text{R} \to N} \mathcal{P}_{\text{M}}+ k_{N \to \text{M}_\text{R}} \mathcal{P}_N, \\[1ex]
\dot{\mathcal{P}}_1 &= - \left(k_{1 \to \text{M}_\text{L}} +k_{1 \to 2}\right) \mathcal{P}_1 \\
&\qquad + k_{{\text{M}_\text{L}} \to 1}\mathcal{P}_{\text{M}} + k_{2 \to 1}\mathcal{P}_2, \\[0ex]
&\ldots& \\
\dot{\mathcal{P}}_a &=  - \left(k_{a \to a+1}+k_{{a \to a-1}}\right) \mathcal{P}_a \\
& \qquad + k_{{a+1 \to a}}\mathcal{P}_{a+1} + k_{{a-1 \to a}}\mathcal{P}_{a-1}, \\[0ex]
&\ldots& \\
\dot{\mathcal{P}}_N &= - \left(k_{N \to \text{M}_\text{R}} + k_{N \to N-1}\right) \mathcal{P}_N \\
& \qquad + k_{{\text{M}_\text{R}} \to N}\mathcal{P}_{\text{M}} + k_{N-1 \to N}\mathcal{P}_{N-1}.
\end{aligned}
\end{equation}
The physical interpretation of this system is that it describes the probability that an electron is localized on each site in the bridge. 
Under steady-state (ss) conditions, $\dot{\mathcal{P}}_a  = 0$ for each state. 
Therefore, when the system is at steady-state the system of differential equations (\ref{eq:occ2}) transforms into a system of algebraic equations.
After augmenting the system with the probability conservation equation 
\begin{equation}
\label{eq:probcons}
\mathcal{P}_{\text{M}} + \sum^N_{a=1} \mathcal{P}_a   = 1,
\end{equation}
the steady-state solution of (\ref{eq:occ2})
can be readily obtained as
 $\boldsymbol{\mathcal{P}}^{(\text{ss})} = \boldsymbol{K}^{-1}\boldsymbol{\mathcal{S}}$ with
$
\boldsymbol{\mathcal{P}}^{(\text{ss})} = \{\mathcal{P}^{(\text{ss})}_{\text{M}},\mathcal{P}^{(\text{ss})}_1,\ldots,\mathcal{P}^{(\text{ss})}_{N}\}
$
where $\mathcal{P}^{(\text{ss})}_a$ is the steady-state probability of state $a$,
$\boldsymbol{K}$ is a matrix of rate coefficients extracted from the LHS of (\ref{eq:occ2}), 
and $\boldsymbol{\mathcal{S}}$ is a column vector with element values that arise from using  Eq.~(\ref{eq:probcons}).
As stated before, we have ignored multiple site occupancies.
The steady-state electronic current
can be evaluated by taking the difference between the unidirectional flux of electrons hopping from left to right and from right to left in the junction, for example,
\begin{equation}
\label{eq:electriccurrent}
\mathcal{J}_\text{el} =  e\left( k_{1 \to \text{M}_\text{L}}\mathcal{P}^{(\text{ss})}_1- k_{\text{M}_\text{L} \to 1} \mathcal{P}^{(\text{ss})}_\text{M}\right).
\end{equation}
In the specific case of $N=1$, which describes a single charge transfer site seated between two electrodes,
the electric current is \cite{Nitzan2011}
\begin{equation}
\label{eq:electriccurrent_1site}
\mathcal{J}_\text{el} = e\left(\frac{k_{1 \to \text{M}_\text{L}}k_{\text{M}_\text{R} \to 1}-k_{1 \to \text{M}_\text{R}}k_{\text{M}_\text{L} \to 1}}{k_{1 \to \text{M}_\text{R}}+k_{\text{M}_\text{R} \to 1}+k_{1 \to \text{M}_\text{R}}+k_{\text{M}_\text{L} \to 1}}\right).
\end{equation}

\subsection{Electron-transfer-induced heat current}

The system of equations that expresses the ETI heat current $\mathcal{J}^{(s)}_\mathcal{Q}$ associated the thermal environment of each site $s$ is (see Fig.~\ref{fig:Main_final}) 
\begin{equation}
\begin{aligned}
\label{eq:heatsystems}
\mathcal{J}^{(1)}_\mathcal{Q} &= -\mathcal{J}^{(\text{L})}_\mathcal{Q} + k_{1 \to 2} \mathcal{P}_1 \mathcal{Q}^{(1)}_{1,2} + k_{2 \to 1} \mathcal{P}_2 \mathcal{Q}^{(1)}_{2,1}, \\
\mathcal{J}^{(2)}_\mathcal{Q} &= k_{1 \to 2}\mathcal{P}_1 \mathcal{Q}^{(2)}_{1,2} + k_{2 \to 1} \mathcal{P}_2 \mathcal{Q}^{(2)}_{2,1} \\
&\qquad + k_{2 \to 3}\mathcal{P}_2 \mathcal{Q}^{(2)}_{2,3} + k_{3 \to 2} \mathcal{P}_3 \mathcal{Q}^{(2)}_{3,2},\\
&\ldots& \\
\mathcal{J}^{(s)}_\mathcal{Q} &= k_{s-1 \to s} \mathcal{P}_{s-1} \mathcal{Q}^{(s)}_{s-1,s} + k_{s \to s-1} \mathcal{P}_{s} \mathcal{Q}^{(s)}_{s,s-1} \\
&\qquad + k_{s \to s+1}\mathcal{P}_s \mathcal{Q}^{(s)}_{s,s+1} + k_{s+1 \to s} \mathcal{P}_{s+1} \mathcal{Q}^{(s)}_{s+1,s},\\[0ex]
&\ldots& \\
\mathcal{J}^{(N)}_\mathcal{Q} &= -\mathcal{J}^{(\text{R})}_\mathcal{Q} + k_{N-1 \to N} \mathcal{P}_{N-1} \mathcal{Q}^{(N)}_{N-1,N} \\
&\qquad  + k_{N \to N-1} \mathcal{P}_N \mathcal{Q}^{(N)}_{N,N-1},
\end{aligned}
\end{equation}
where $\mathcal{J}^{(\text{L})}_\mathcal{Q}$ and $\mathcal{J}^{(\text{R})}_\mathcal{Q}$ are the molecule-to-metal ETI heat currents at the left and right electrodes,
$k_{a \to b} \mathcal{P}_a$ is a unidirectional electron flux term,
and \cite{craven16c}
\begin{equation}
\label{eq:heat}
\mathcal{Q}^{(s)}_{a,b} = \frac{ \displaystyle T_s \lambda^{(a,b)}_{s} \Delta E_{ab} + \left(T_\alpha - T_s\right) \lambda^{(a,b)}_{a} \lambda^{(a,b)}_{b}  }{\displaystyle  T_a \lambda^{(a,b)}_{a}+  T_b \lambda^{(a,b)}_{b} },
\end{equation}
is the  heat change in the thermal environment of site $s$ during the electronic transition $a \to b$ with $\alpha$ being the element in $\left\{a,b\right\}$ that is not equal to $s$. 
Note that Eq.~(\ref{eq:heat}) is the heat change \textit{per electron}.
We are interested in the steady-state limit of the system of equations (\ref{eq:heatsystems}) which is obtained by replacing all the $\mathcal{P}_a$ terms with the corresponding $\mathcal{P}^{(\text{ss})}_a$ terms in each heat current expression.

The molecule-to-metal heat current expressions can be constructed by examining how ET across the thermal gradient between the terminal molecules in the bridge (1 or $N$) and the corresponding electrode affects heat transport. \cite{craven17b} 
Using the approximation that there is only a single vibrational mode associated with each molecular site, 
ET at the left electrode only affects mode $x_1$ and ET at the right electrode only affects mode $x_N$.
Therefore, the ETI heat current between molecule and metal at the left electrode is
\begin{equation}
\begin{aligned}
\label{eq:heatcurrentAbeL}
\mathcal{J}^{(\text{L})}_\mathcal{Q}
&=   \int_{-\infty}^{\infty} \mathcal{P}_1 \big[1-f\big(T_\text{L}, \mu_\text{L}, E_{\text{L}}(x_1)\big)\big] \Gamma \big(E_{\text{L}}(x_1)\big) \\
& \quad \times \big( E_{\text{L}}(x_1)-\mu_\text{L} \big) \frac{\exp\big[- E_{1,1}^\ddag(x_1)  \big/  k_\text{B} T_1 \big]}{Z_{1,1}^\ddag}\,dx_1\\
&\quad +\int_{-\infty}^{\infty} \mathcal{P}_\text{M} \, f\big(T_\text{L}, \mu_\text{L}, E_{\text{L}}(x_1)\big)\Gamma \big(E_{\text{L}}(x_1)\big) \\
& \quad \times \big(\mu_\text{L}  - E_{\text{L}}(x_1)\big) \frac{\exp\big[- E_{\text{M},1}^\ddag(x_1) \big/ k_\text{B} T_1 \big]}{Z_{\text{M},1}^\ddag}\,dx_1, \\ 
\end{aligned}
\end{equation}
and at the right electrode 
\begin{equation}
\begin{aligned}
\label{eq:heatcurrentAbeR}
\mathcal{J}^{(\text{R})}_\mathcal{Q}
&=   \int_{-\infty}^{\infty}\mathcal{P}_N \big[1-f\big(T_\text{R}, \mu_\text{R}, E_{\text{R}}(x_N)\big)\big] \Gamma_\text{R} \big(E_{\text{R}}(x_N)\big) \\
& \quad \times \big( E_{\text{R}}(x_N)-\mu_\text{R} \big) \frac{\exp\big[- E_{N,N}^\ddag(x_N)\big/ k_\text{B} T_N\big]}{Z_{N,N}^\ddag}\,dx_N\\
&\quad +\int_{-\infty}^{\infty} \mathcal{P}_\text{M} \,  f\big(T_\text{R}, \mu_\text{R}, E_{\text{R}}(x_N)\big)\Gamma_\text{R} \big(E_{\text{R}}(x_N)\big) \\
& \quad \times \big(\mu_\text{R} - E_{\text{R}}(x_N)\big) \frac{\exp\big[- E_{\text{M},N}^\ddag(x_N) \big/ k_\text{B} T_N \big]}{Z_{\text{M},N}^\ddag}\,dx_N, \\ 
\end{aligned}
\end{equation}
where 
\begin{equation}
\label{eq:act}
E_{a,j}^\ddag(x_s) = \frac{1}{2}k_j\left(x_s - \bar{x}^{(a)}_s\right)^2, \\
\end{equation}
is the energy of mode $x_s$ above the corresponding energy origin $E'_a$ when the system is in state $a$,
\begin{align}
\label{eq:EabxL}
\nonumber E_\text{L}(x_1) &= k_1\Big(\bar{x}^{(\text{M})}_1 -\bar{x}^{(1)}_1 \Big) x_1 +\frac{1}{2} k_1 \left(\bar{x}^{(1)}_1\right)^2 \\
&\quad -\frac{1}{2} k_1 \left(\bar{x}^{(\text{M})}_1\right)^2- \Delta E_\text{L},
\end{align}
is the energy difference between surfaces $E_1$ and $E_\text{M}$ in mode $x_1$, 
\begin{align}
\label{eq:EabxR}
\nonumber E_\text{R}(x_N) &= k_N\Big(\bar{x}^{(\text{M})}_N -\bar{x}^{(N)}_N \Big) x_N +\frac{1}{2} k_N \left(\bar{x}^{(N)}_N\right)^2 \\
&\quad - \frac{1}{2} k_N \left(\bar{x}^{(\text{M})}_N\right)^2 - \Delta E_\text{R}, 
\end{align}
is the energy difference between surfaces $E_N$ and $E_\text{M}$ in mode $x_N$, and
\begin{equation}
\label{eq:part}
Z_{a,s}^\ddag = \int_{-\infty}^{\infty} \exp\big[-E_{a,s}^\ddag(x_s)/k_\text{B} T_s\big] \,dx_s = \sqrt{\frac{2 \pi k_\text{B} T_s}{k_s}},
\end{equation}
is a normalization factor.
The steady-state limit of $\mathcal{J}^{(\text{L})}_\mathcal{Q}$ and $\mathcal{J}^{(\text{R})}_\mathcal{Q}$ are obtained 
by replacing the $\mathcal{P}_1$ and $\mathcal{P}_N$ terms with $\mathcal{P}^{(\text{ss})}_1$ and $\mathcal{P}^{(\text{ss})}_N$, respectively.

\subsection{\label{nummethods} Numerical implementation and physical approximations}

Several specific numerical procedures are employed to calculate and analyze ETI thermal conductance:

(a)  A variant of the self-consistent reservoir (SCR) method \cite{Bolsterli1970SCR,Bonetto2004SCR,Segal2009SCR,Tulkki2013SCR} is applied to define the temperature profile in the molecular bridge.
In this modified SCR procedure, the bridge temperatures $T_1, T_2, \ldots, T_N$ and the voltage bias $V$ are obtained subject to the constraints:
\begin{equation}
\label{eq:constraint}
\mathcal{J}^{(s)}_\mathcal{Q} = 0 \, \forall \, s \in \left\{1,2,\ldots,N\right\} \quad \text{and} \quad \mathcal{J}_\text{el} = 0.
\end{equation}
These constraints state that the heat current into/out of the thermal bath associated with each molecular site and also the electronic current through the junction must vanish.
This is a model that could describe an isolated molecular bridge where the nuclear environment is intramolecular or the situation in which there is a temperature gradient in the solvent environment in a solvated molecular junction.
The latter constraint in Eq.~(\ref{eq:constraint}) does not appear in the traditional SCR method that is normally applied to examine phononic heat transport mechanisms.
A multidimensional Newton-Raphson root-finding procedure is used to satisfy (\ref{eq:constraint}).
Using the SCR procedure ensures that there is no \textit{net} heat flow associated with the thermal environment of each molecular site and therefore the magnitude of the heat current through the junction $\mathcal{J}_\mathcal{Q}$ can be well defined as
\begin{equation}
\mathcal{J}_\mathcal{Q} = \big|\mathcal{J}^{(\text{L})}_\mathcal{Q}\big| = \big|\mathcal{J}^{(\text{R})}_\mathcal{Q}\big|.
\end{equation}

(b)  The expression for the thermal conductance on the RHS of Eq.~(\ref{eq:themalcond}) is evaluated by calculating the heat current at a specific temperature $T$ with $\Delta T = \Delta T_+ = 1\,\text{K}$ and again with $\Delta T = \Delta T_- = -1\,\text{K}$, and then reporting the measured thermal conductance as
\begin{equation}
\label{eq:kappa}
\mathcal{K} = \frac{\mathcal{J}_\mathcal{Q}(\Delta T_+) - \mathcal{J}_\mathcal{Q}(\Delta T_-)}{\Delta T_+-\Delta T_-}.
\end{equation}
All calculations are performed under zero electric current conditions.

(c)  Four energy landscapes are used: 
$\mathcal{L}_1 \equiv E'_s = 0 \, \forall \, s$, 
$\mathcal{L}_2 \equiv E'_s = 0.05\,\text{eV} \, \forall \, s$, 
$\mathcal{L}_3$ which is linear ramp of energy levels from $0.05\,\text{eV}$ to  $-0.05\,\text{eV}$,
and 
$\mathcal{L}_4 \equiv E'_s = 0.1\,\text{eV} \, \forall \, s$.
Note that at least two sites are needed in a molecular bridge to construct the energy gradient in $\mathcal{L}_3$.

We also apply several physical approximations in our model:

(a) The wide-band approximation is applied in all numerical calculations and thus $\Gamma_\text{L}$ and $\Gamma_\text{R}$ are taken to be independent of the energy level $\epsilon$. 
Furthermore, throughout, we take the coupling at the left and right electrodes to be equal $\Gamma_\text{L} = \Gamma_\text{R} = \Gamma$.

(b) We also assume that electron hopping between sites only affects the two sites and the associated vibrational environments involved in that process. 
This implies that the total reorganization energy for the transition $a \to b$ is given by $\lambda^{(a,b)} = \lambda^{(a,b)}_{a} + \lambda^{(a,b)}_{b}$. 
We take the site reorganization energies to be equal $\lambda^{(a,b)}_{a} = \lambda^{(a,b)}_{b}$. 
The total reorganization energy is assumed to be the same for each state transition, and we denote this reorganization energy as $\lambda$.

(c) We assume that the nuclear system is not transferring energy to the electrodes directly by a phononic mechanism.

\section{Ballistic Electron Transport \label{sec:ballistic}}

The ETI heat current can be compared with the Landauer electronic heat current to illustrate how hopping ET and ballistic electron transport respectively manifest in the thermal transport properties of the junction.
The ballistic currents are evaluated using Landauer theory and a nonequilibrium Green's function approach.

In the Landauer limit of transport, electronic motion through the molecular bridge is ballistic and electron-phonon interaction is disregarded.
This is the opposite limit of transport with respect to electron-phonon coupling strength than the electron hopping mechanism examined in ETIHT.
The ballistic electric current through the junction can be expressed using the Landauer formalism as \cite{Galperin2008}
\begin{equation}
\mathcal{J}_\text{el} = \frac{e}{\pi \hbar}\int_{-\infty}^{\infty}\!\!\mathcal{T}(\epsilon)\Big[ f(T_\text{R},\mu_\text{R},\epsilon) - f(T_\text{L},\mu_\text{L},\epsilon)  \Big] d\epsilon, 
\end{equation}
where $\mathcal{T}(\epsilon)$ is the electronic transmission function.
The transmission function is constructed using a nonequilibrium Green's function approach as
\begin{equation}
\label{eq:Trans}
\mathcal{T}(\epsilon) = \text{Tr}\left[\hat{\Gamma}_\text{L}(\epsilon) \hat{G}^\dag(\epsilon) \hat{\Gamma}_\text{R}(\epsilon) \hat{G}(\epsilon)\right],
\end{equation}
where $\hat{G}$ is the Green's function for the molecular bridge that satisfies
\begin{widetext}
\renewcommand*{\arraystretch}{1.2}
\begin{equation}
			\hat{G}^{-1}(\epsilon)=
		\begin{pmatrix}
			\epsilon - E'_1 + \frac{1}{2} i \hbar  \Gamma_\text{L}(\epsilon)& V_{1,2}&0&0&\ldots&0 \\
		V_{2,1}&\epsilon - E'_2&V_{2,3}&0&\ldots&0 \\
			0&V_{3,2}&\epsilon - E'_3&V_{3,4}& & \vdots\\
			0&0&V_{4,3}&\ddots&\ddots & 0\\
		\vdots&\vdots& &\ddots&\epsilon-E'_{N-1} & V_{N-1,N} \\
			0&0&\cdots&0&V_{N,N-1} & \epsilon - E'_N + \frac{1}{2} i \hbar  \Gamma_\text{R}(\epsilon)\\
		\end{pmatrix}.
\end{equation}
\end{widetext}
The molecule-metal coupling matrices
\begin{equation}
 			\hat{\Gamma}_\text{L}(\epsilon) =
		\begin{pmatrix}
			\hbar \Gamma_\text{L}(\epsilon)&0&\cdots&0 \\
			0&0&\cdots&0 \\
			\vdots&\vdots&\ddots&\vdots \\
			0&0&\cdots&0 \\
			\end{pmatrix},
\end{equation}
and			
\begin{equation}				
			\hat{\Gamma}_\text{R}(\epsilon) =
		\begin{pmatrix}
			0&0&\cdots&0 \\
			0&0&\cdots&0 \\
			\vdots&\vdots&\ddots&\vdots \\
			0&0&\cdots&\hbar \Gamma_\text{R}(\epsilon) \\
		\end{pmatrix},
\end{equation}
define the interaction between the respective terminal molecular site ($1$ or $N$) at each end of the bridge and the corresponding electrode ($\text{M}_\text{L}$ or $\text{M}_\text{R}$).
We again apply the wide-band approximation implying that $\Gamma_\text{L}$ and $\Gamma_\text{R}$ are taken to be constant and equal ($\Gamma_\text{L} = \Gamma_\text{R} = \Gamma$) in all numerical calculations.
Ballistic electron transport through the junction generates an electronic heat current that can be calculated using the Landauer expression \cite{Cui2017perspective}
\begin{equation}
\label{eq:heatcurrentappendix}
\mathcal{J}_\mathcal{Q} = \frac{1}{\pi \hbar}\int_{-\infty}^{\infty}\!\!\mathcal{T}(\epsilon)\left(\epsilon - \mu\right)  \Big[ f(T_\text{R},\mu_\text{R},\epsilon) - f(T_\text{L},\mu_\text{L},\epsilon)  \Big] d\epsilon,
\end{equation}
which is evaluated under the condition of zero electronic current $\mathcal{J}_\mathcal{Q}|_{\mathcal{J}_\text{el}=0}$. 
The zero electronic current condition is enforced by varying the voltage bias $V$. 
The electronic thermal conductance generated by ballistic electron transport is calculated numerically by combining Eq.~(\ref{eq:heatcurrentappendix}) with Eq.~(\ref{eq:kappa}). 

\section{\label{sec:results}Results}

\begin{figure}
\includegraphics[width = 8.5cm,clip]{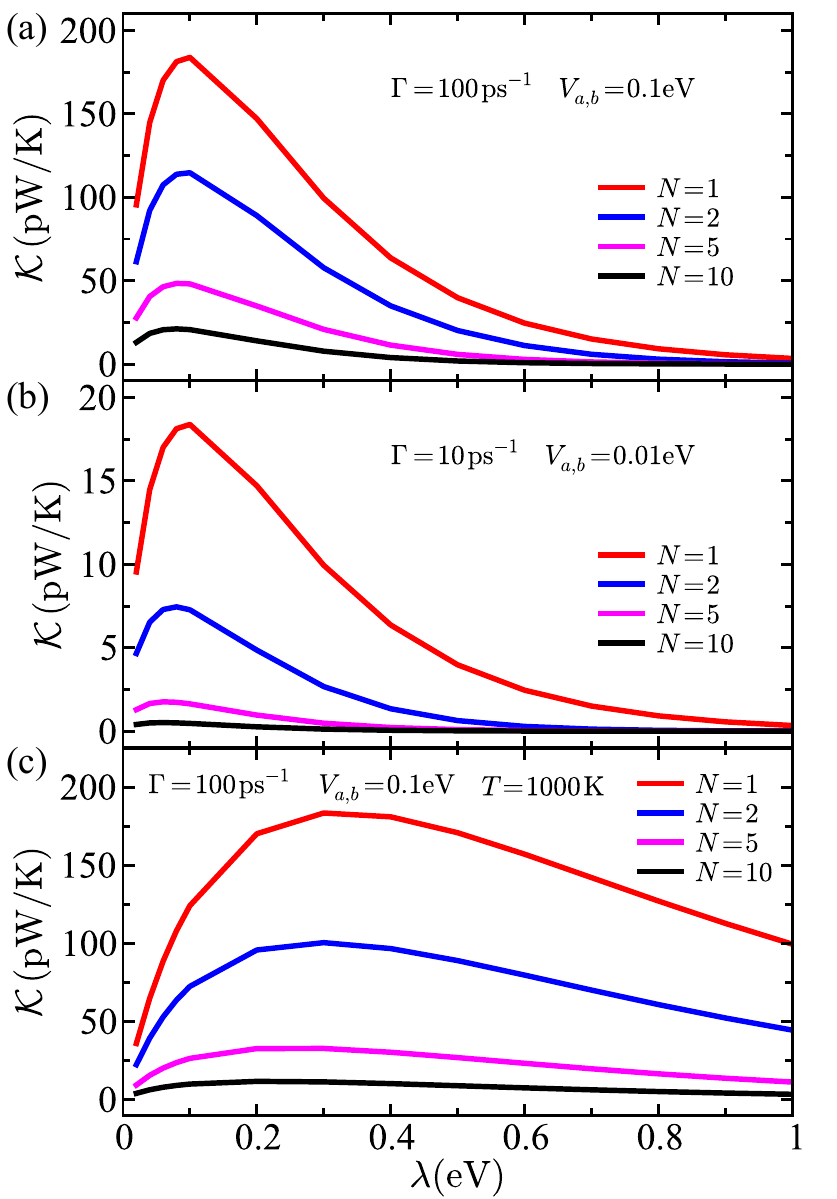}
\caption{\label{fig:ER1}
Thermal conductance $\mathcal{K}$ as a function of total reorganization energy $\lambda$ for bridges with
$N=1$ (red), 
$N=2$ (blue), 
$N=5$ (magenta), 
and
$N=10$ (black) sites.
The electronic coupling parameters are $\Gamma = 100\,\text{ps}^{-1}$ and $V_{a,b} = 0.1\,\text{eV}$ in (a) and (c) and $\Gamma = 10\,\text{ps}^{-1}$ and $V_{a,b} = 0.01\,\text{eV}$ in (b).
The energy landscape in all calculations is $\mathcal{L}_1 \equiv E'_s = 0 \, \forall \, s$.
The temperature is $300\,\text{K}$ in (a) and (b) and $1000\,\text{K}$ in (c).
Other parameters are $\mu = 0$.
}
\end{figure}

The ETI thermal conductance values are shown in Fig.~\ref{fig:ER1} as function of reorganization energy $\lambda$. 
Results are shown for various values of $N$.
The electronic coupling parameters are $\Gamma = 100\,\text{ps}^{-1}$ and $V_{a,b} = 0.1\,\text{eV}$ in Fig.~\ref{fig:ER1}(a) and (c) and $\Gamma = 10\,\text{ps}^{-1}$ and $V_{a,b} = 0.01\,\text{eV}$ in Fig.~\ref{fig:ER1}(b).
These electronic couplings are taken from previously estimated values. \cite{Blumberger2014,Blumberger2014ultrafast,Blumberger2016,Blumberger2017metal,Blumberger2018organic}
It is important to note that our choice of parameter values is designed, in part, to probe ETHIT at the limits of validity of Marcus kinetics. The coupling parameter values $\Gamma = 100\,\text{ps}^{-1}$ and $V_{a,b} = 0.1\,\text{eV}$ can be considered a limiting case for Marcus theory, in range of the transition from nonadiabatic Marcus hopping to strong coupling and/or adiabatic ET mechanisms using the temperatures and reorganization energies examined here.\cite{Jain2015}
At a temperature of $300\,\text{K}$, shown in Fig.~\ref{fig:ER1}(a) and (b), the thermal conductance is maximized at $\lambda \approx 0.1\,\text{eV}$.
Comparing Fig.~\ref{fig:ER1}(a) and Fig.~\ref{fig:ER1}(b), it can be inferred that the shape of the thermal conductance curve with respect to variation of $\lambda$ does not depend strongly on the electronic coupling parameters, 
however its magnitude does depend on these parameters.
The observation that thermal conductance tends to zero in the $\lambda \to 0$ and $\lambda \to \infty$ limits and goes through a single maximum between those limits agrees with the observations in Ref.~\citenum{craven21a}.
The cause of this behavior is that the ETI thermal conductance curves follow the electron transfer rates, which also exhibit this behavior.
In Fig.~\ref{fig:ER1}(c) we use the same electronic properties as in Fig.~\ref{fig:ER1}(a), but with the temperature increased to $1000\,\text{K}$. The same general trends are observed, however, the maximum conductance is reached at $\lambda \approx 0.3\,\text{eV}$ as opposed the maximum at $\lambda \approx 0.1\,\text{eV}$ observed for 300K. 
This thermal conductance behavior again follows the functional behavior of the electron transfer rates, i.e., the maximum thermal conductance is obtained when the electron transfer rates are maximized. 

The magnitude of the calculated ETI thermal conductances are notable.  
In Fig.~\ref{fig:ER1}(a), the maximum conductance for $N=1$ is $\approx175\,\text{pW/K}$, a value that is approximately $10$ times greater than the $\approx20\,\text{pW/K}$ value that has recently been measured in alkane molecular junctions using single-molecule experiments. \cite{Reddy2019nature,Mosso2019}
For $N=2$, the maximum conductance is $\approx 120\,\text{pW/K}$, also a significant value compared to the measured single-molecule phononic thermal conductance values.
Figure~\ref{fig:ER1}(b) shows results with both the molecule-metal and molecule-molecule electronic couplings reduced a factor of $10$.
Making the electronic couplings weaker significantly reduces the magnitude of the ETI thermal conductance, however for molecular bridges of length $N=1$ and $N=2$, the maximum thermal conductance is still of the same order to the measured single-molecule values that are mostly attributed to phononic heat transport. 
It is notable that increasing the system temperature to $1000\,\text{K}$, as shown in Figure~\ref{fig:ER1}(c), changes the functional shape of the thermal conductance curves but does not significantly change their maximum value.
There are three important implications of these predicted magnitudes: 
(a) the ETI thermal conductances are in range of what is currently experimentally measurable,
(b) the ETI thermal conductance could make an appreciable contribution to the overall thermal conductance and, 
(c) ETIHT could be the principle heat transport mechanism in properly engineered systems.

\begin{figure}
\includegraphics[width = 8.5cm,clip]{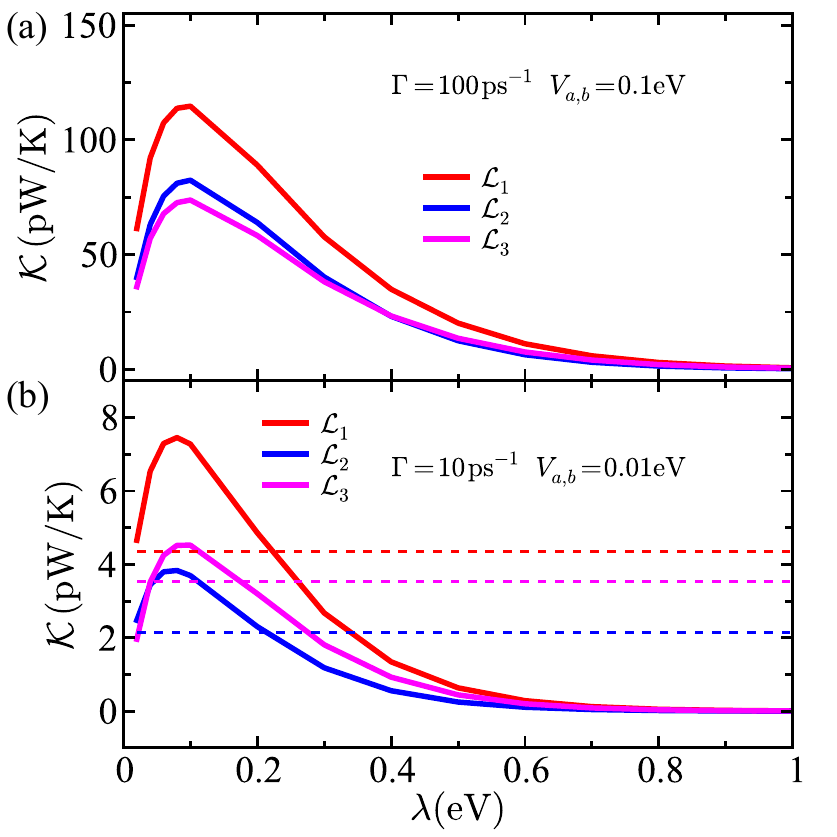}
\caption{\label{fig:ER2}
Thermal conductance $\mathcal{K}$ as a function of total reorganization energy $\lambda$ for energy landscapes 
$\mathcal{L}_1 \equiv E'_s = 0 \, \forall \, s$ (red), 
$\mathcal{L}_2 \equiv E'_s = 0.05\,\text{eV} \, \forall \, s$ (blue), 
and $\mathcal{L}_3$ (magenta) which is linear ramp of energy levels from $0.05\,\text{eV}$ to  $-0.05\,\text{eV}$.
Parameters are (a) $N=2$, $\Gamma = 100\,\text{ps}^{-1}$ and $V_{a,b} = 0.1\,\text{eV}$ and (b) $N=2$, $\Gamma = 10\,\text{ps}^{-1}$ and $V_{a,b} = 0.01\,\text{eV}$.
Other parameters in both panels are $\mu = 0$ and $T = 300\,\text{K}$.
The dashed lines in (b) correspond to the values obtained in the ballistic electron transport limit using Landauer theory.}
\end{figure}

Figure~\ref{fig:ER2} illustrates how variations in the energy landscape affect the thermal conductance for different values of the reorganization energy. 
The red curves in both Figure~\ref{fig:ER2} (a) and (b) correspond to $\mathcal{L}_1 \equiv E'_s = 0 \, \forall \, s$, the  blue curves correspond to $\mathcal{L}_2 \equiv E'_s = 0.05\,\text{eV} \, \forall \, s$  and the magenta curves correspond to $\mathcal{L}_3$ which is linear ramp of energy levels from $0.05\,\text{eV}$ to  $-0.05\,\text{eV}$. 
The $\mathcal{L}_1$ landscape represents the situation in which all of the electronic occupation energies in the molecular bridge are aligned with the unbiased Fermi levels of the electrodes.
In Fig.~\ref{fig:ER2}(a) the electronic coupling parameters are $\Gamma = 100\,\text{ps}^{-1}$, $V_{a,b} = 0.1\,\text{eV}$ and in Fig.~\ref{fig:ER2}(b) they are $\Gamma = 10\,\text{ps}^{-1}$, $V_{a,b} = 0.01\,\text{eV}$.
Similar trends are observed for the different landscapes. 
For each $\mathcal{L}$, the thermal conductance tends to zero for $\lambda \to 0$ and $\lambda \to \infty$ and goes through a single maximum between these limits, following the same behavior in the molecule-metal electron transfer rates.
The $\lambda$ value at which $\mathcal{K}$ is maximized is different for each landscape, again reflecting the same behavior in the electron transfer rates. 
The dashed lines in Fig.~\ref{fig:ER2}(b) represent the values obtained in the ballistic electron transport limit using Landauer theory for the respective energy landscape. 
The primary observation that can be taken from these ballistic calculations is that the ETI thermal conductance can exceed the ballistic thermal conductance in certain regimes of electronic coupling strength. 
In general, however, electron-phonon coupling that manifests through the electron transfer reorganization energy reduces the thermal transport resulting in lower values for the thermal conductance---the examples shown Fig.~\ref{fig:ER2}(b) are an exception.

\begin{figure}
\includegraphics[width = 8.5cm,clip]{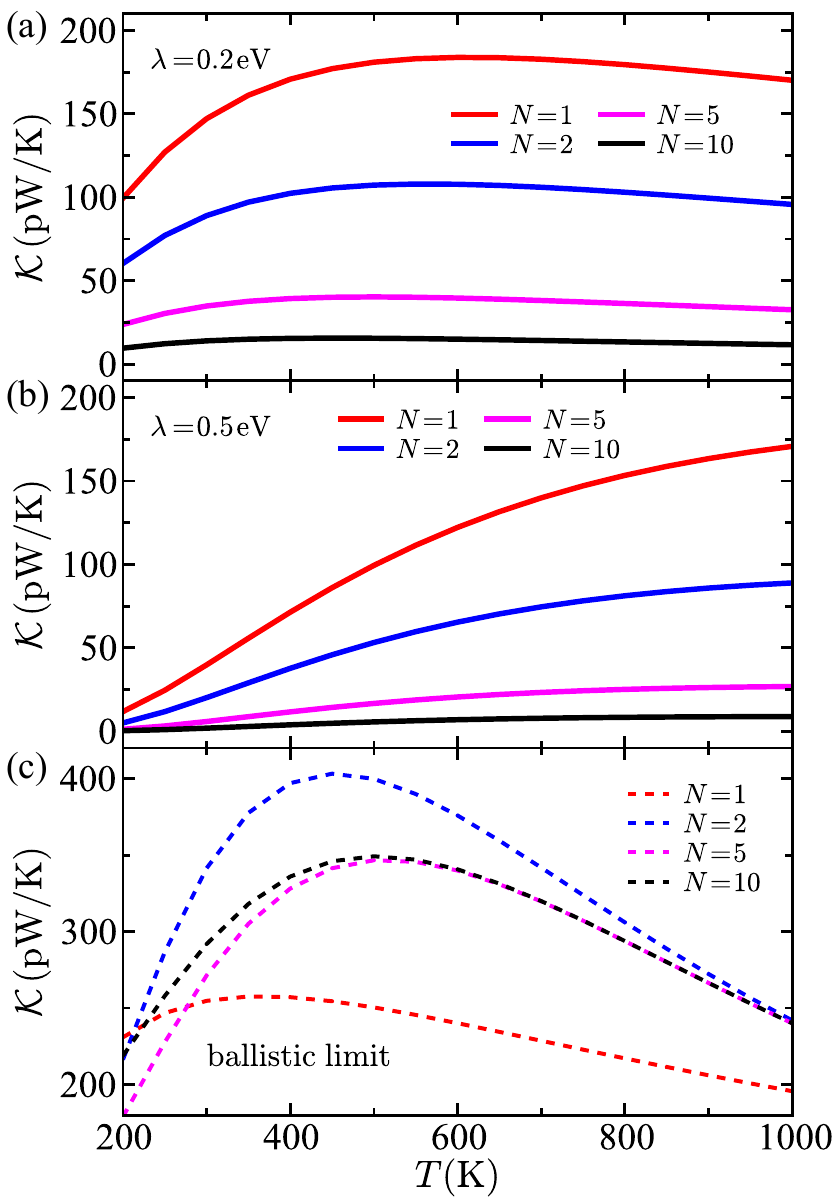}
\caption{\label{fig:T}
Thermal conductance $\mathcal{K}$ as a function of temperature $T$  for bridges with
$N=1$ (red), 
$N=2$ (blue), 
$N=5$ (magenta), 
and
$N=10$ (black) sites.
Parameters are (a) $\lambda = 0.2\,\text{eV}$ and (b) $\lambda = 0.5\,\text{eV}$.
Other parameters are $\Gamma = 100\,\text{ps}^{-1}$, $V_{a,b} = 0.1\,\text{eV}$, and $\mu = 0$.
The bottom panel (c) shows the results calculated in the ballistic electron transport limit using Landauer theory with the same parameter values.
The dashed curves in (c) denote that the calculations are performed in the ballistic limit.
The energy landscape in all calculations is $\mathcal{L}_1$.
}
\end{figure}

The temperature dependence of the thermal conductance is shown in Fig.~\ref{fig:T}. 
Results are shown for molecular bridges of various lengths in each panel.
In all cases, the ETI thermal conductance is greater the shorter the molecular chain. 
It is notable that the thermal conductance does not increase linearly and without bound as the temperature is increased.
Instead, a maximum value is reached when $T \approx \lambda/ 4 k_\text{B}$ for the $\mathcal{L}_1$ landscape examined here.
This maximum thermal conductance temperature coincides with temperature that maximizes the electron transfer rates. 
The ballistic limit results are shown in Fig.~\ref{fig:T}(c). 
The first point to note is that the thermal conductance due to ballistic transport is typically greater than the thermal conductance due to ETI transport. 
This implies that, in general, the electron-phonon coupling, here expressed through the reorganization energy, will reduce the amount of heat that is transported. 
This is particularly pronounced for long chains (large values of $N$).
Also note that for ballistic transport the thermal conductance is not a monotonically decreasing function with increasing $N$. 
This can be observed, for example, by noting that the conductance values $N=2$ are greater than $N=1$ values. 
This behavior is observed for ballistic transport and not for ETI transport, an interesting distinguishing characteristic between the two mechanisms.

\begin{figure}
\includegraphics[width = 8.5cm,clip]{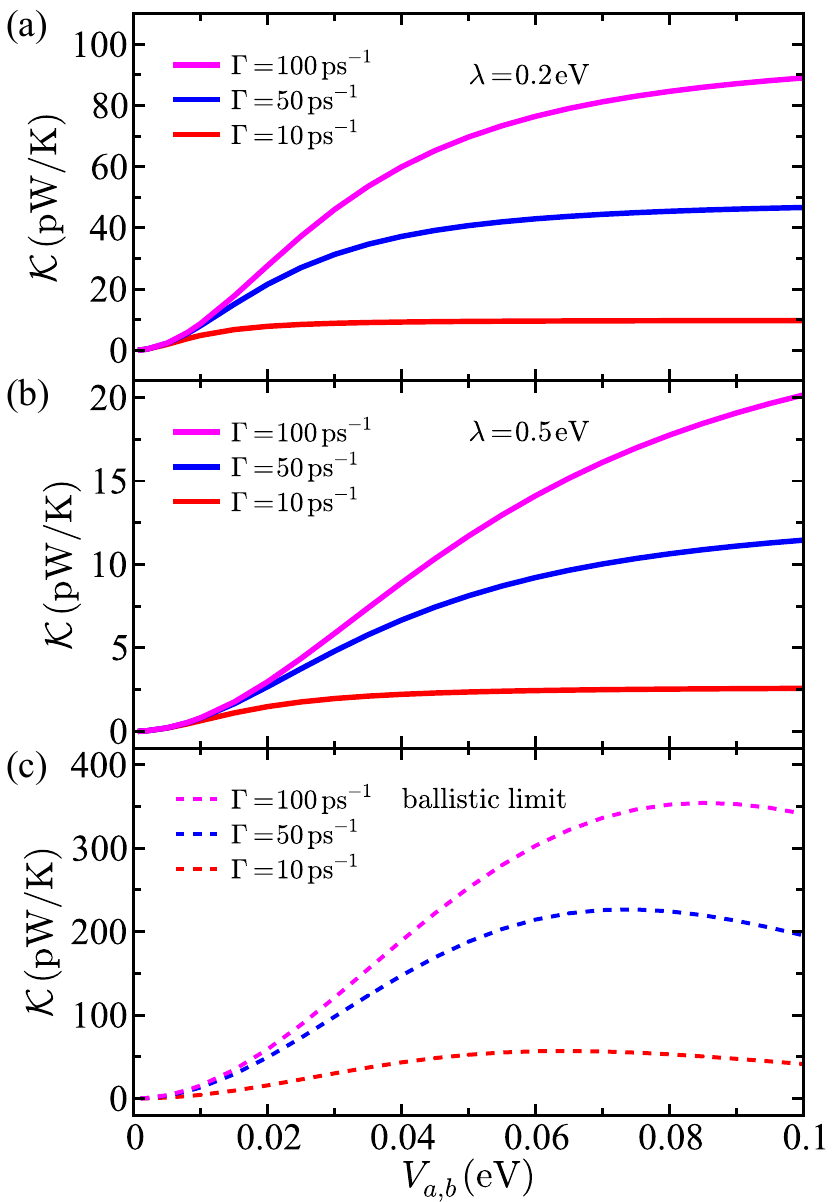}
\caption{\label{fig:V}
Thermal conductance $\mathcal{K}$ as a function of electronic coupling $V_{a,b}$ for bridges with
$\Gamma = 10\,\text{ps}^{-1}$ (red),
$\Gamma = 50\,\text{ps}^{-1}$ (blue), 
and $\Gamma = 100\,\text{ps}^{-1}$ (magenta).
Parameters are (a) $\lambda = 0.2\,\text{eV}$ and (b) $\lambda = 0.5\,\text{eV}$.
Other parameters are $N = 2$, $T = 300\,\text{K}$, and $\mu = 0$.
The bottom panel (c) shows the results calculated in ballistic electron transport limit using Landauer theory with the same parameter values.
The dashed curves in (c) denote that the calculations are performed in the ballistic limit.
The energy landscape in all calculations is $\mathcal{L}_1$.
}
\end{figure}

The dependence of the thermal conductance on electronic coupling between molecular sites $V_{a,b}$ is shown in  Fig.~\ref{fig:V}. 
A molecule bridge with $N = 2$ sites is used in all calculations. 
In the limiting case of $V_{a,b} \to 0$ we observe the expected behavior that the thermal conductance vanishes. 
This simply implies the intuitive result that when the two sites in the bridge are decoupled, thermal conductance due to electronic motion goes to zero. 
In each plot, results are shown for 
$\Gamma = 10\,\text{ps}^{-1}$ (red),
$\Gamma = 50\,\text{ps}^{-1}$ (blue), 
and $\Gamma = 100\,\text{ps}^{-1}$ (magenta).
Figure~\ref{fig:V}(a) shows the results for $\lambda = 0.2\,\text{eV}$. 
The thermal conductance exhibits a nonlinear dependence on $V_{a,b}$, with the results following functional forms similar to logistic curves.
Figure~\ref{fig:V}(b) shows results for $\lambda = 0.5\,\text{eV}$. 
The thermal conductance again exhibits nonlinear dependence on $V_{a,b}$. 
The corresponding ballistic thermal conductance values are shown in Fig.~\ref{fig:V}(c).
Comparing the ETI and ballistic results, there are two important observations: 
(a) the ballistic results goes through a maximum and then decreases as $V_{a,b}$ is increased, behavior that is not observed for the ETI results
and (b) the maximum ballistic thermal conductance value is larger than the maximum ETI value.

\begin{figure}
\includegraphics[width = 8.5cm,clip]{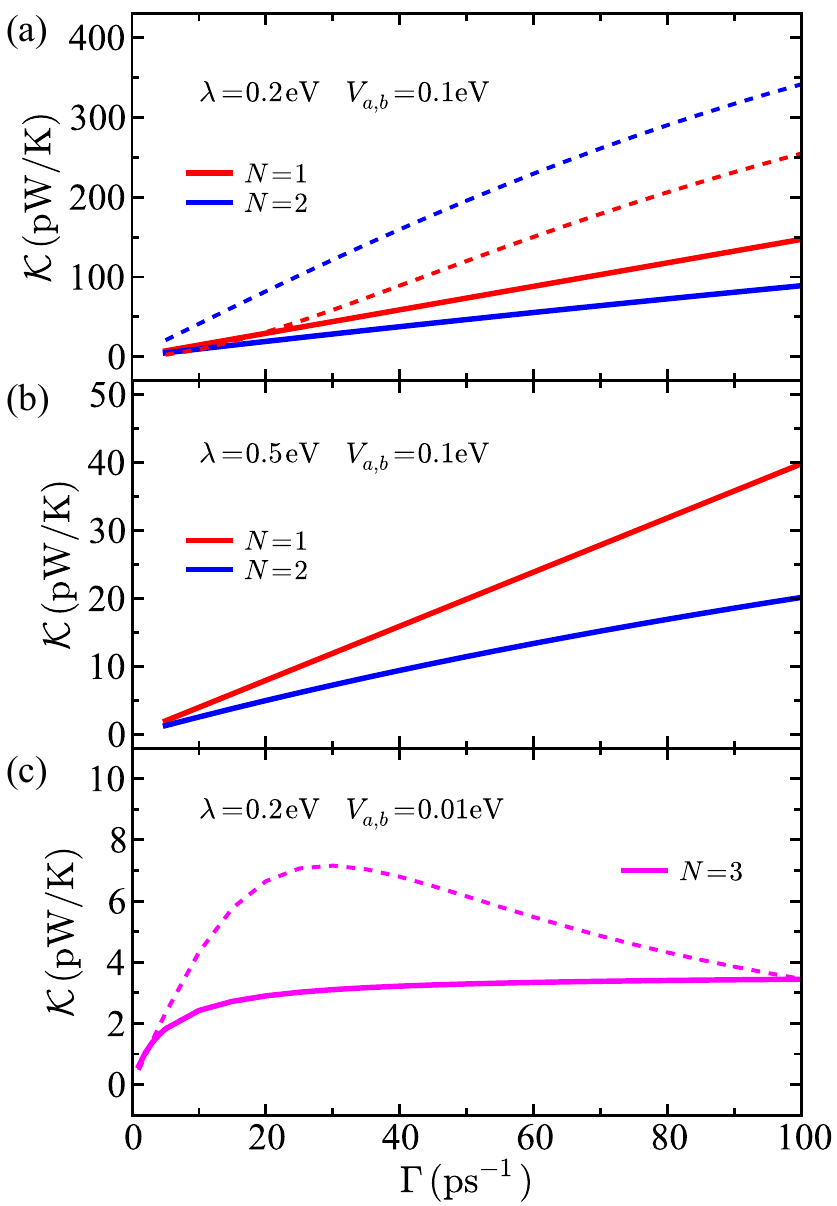}
\caption{\label{fig:Gamma}
Thermal conductance $\mathcal{K}$ as a function of molecule-metal electronic coupling strength $\Gamma$.
Parameters are (a) $N=1$ (red) and $N=2$ (blue), and $\lambda = 0.2\,\text{eV}$ and (b) $\lambda = 0.5\,\text{eV}$.
Other parameters are $T = 300\,\text{K}$, $V_{a,b} = 0.1\,\text{eV}$, and $\mu = 0$.
The bottom panel (c) shows the results calculated for an in the ballistic electron transport limit using Landauer theory with the same parameter values.
In all panels, the solid curves denote results calculated in the hopping limit and the dashed curves denote the results calculated in the ballistic electron transport limit using Landauer theory with the same parameter values.
The energy landscape in all calculations is $\mathcal{L}_1$.
}
\end{figure}

The dependence of the thermal conductance on molecule-metal coupling strength $\Gamma$ is shown in  Fig.~\ref{fig:Gamma}. Analogous to the the molecule-molecule electronic coupling case, in the limit $\Gamma \to 0$ the thermal conductance vanishes due to the molecular bridge being decoupled from the metal electrodes in this limit.
In Fig.~\ref{fig:Gamma}(a) and (b), curves are shown for $N = 1$ (red) and $N = 2$ (blue) sites. 
Note that for $N=1$, the thermal conductance does not depend on $V_{a,b}$.
In Fig.~\ref{fig:Gamma}(a), with $\lambda = 0.2\,\text{eV}$, the ETI thermal conductance scales approximately linearly with $\Gamma$ and reaches a maximum value of $\sim 100 \,\text{pW/K}$ at the end of the examined range. The corresponding ballistic results are shown by dashed curves. 
The magnitude of the ballistic values for $\mathcal{K}$ are in general greater than the ETI values.
Figure~\ref{fig:Gamma}(b) shows the results with the reorganization energy increased to $\lambda = 0.5\,\text{eV}$.
The scaling is again approximately linear in $\Gamma$ and reaches maximum values of $\approx 40 \,\text{pW/K}$ for $N=1$ and $\approx 20 \,\text{pW/K}$ for $N=2$. 
The results for a bridge with $N=3$ sites are shown in Fig.~\ref{fig:Gamma}(c) with $V_{a,b}$ reduced by a factor 10 to $0.01\,\text{eV}$. In this case,
the thermal conductance rises sharply for small $\Gamma$, but then shows an approximately linear increase. 
The corresponding ballistic result (dashed curve) goes through a maximum where its magnitude is approximately a factor $2$ greater than
ETI thermal conductance, but then decays with increasing $\Gamma$.


Fourier's law predicts that heat current in molecule or molecular junction with $N$ sites will scale as
\begin{equation}
\mathcal{J}_\mathcal{Q} \propto \frac{\Delta T}{N},
\end{equation}
i.e, that the heat current $\mathcal{J}_\mathcal{Q} \propto N^{-1}$.
In rest of this Section, we examine under what conditions the ETI thermal conductance obeys Fourier's law and under what conditions deviations from Fourier’s law are observed.
To do so, we apply the the general scaling formula
\begin{equation}
\mathcal{J}_\mathcal{Q} \propto N^\alpha
\end{equation}
and extract values of $\alpha$ for various physical conditions.
When $\alpha = -1$ Fourier's law behavior is observed.

\begin{figure}
\includegraphics[width = 8.5cm,clip]{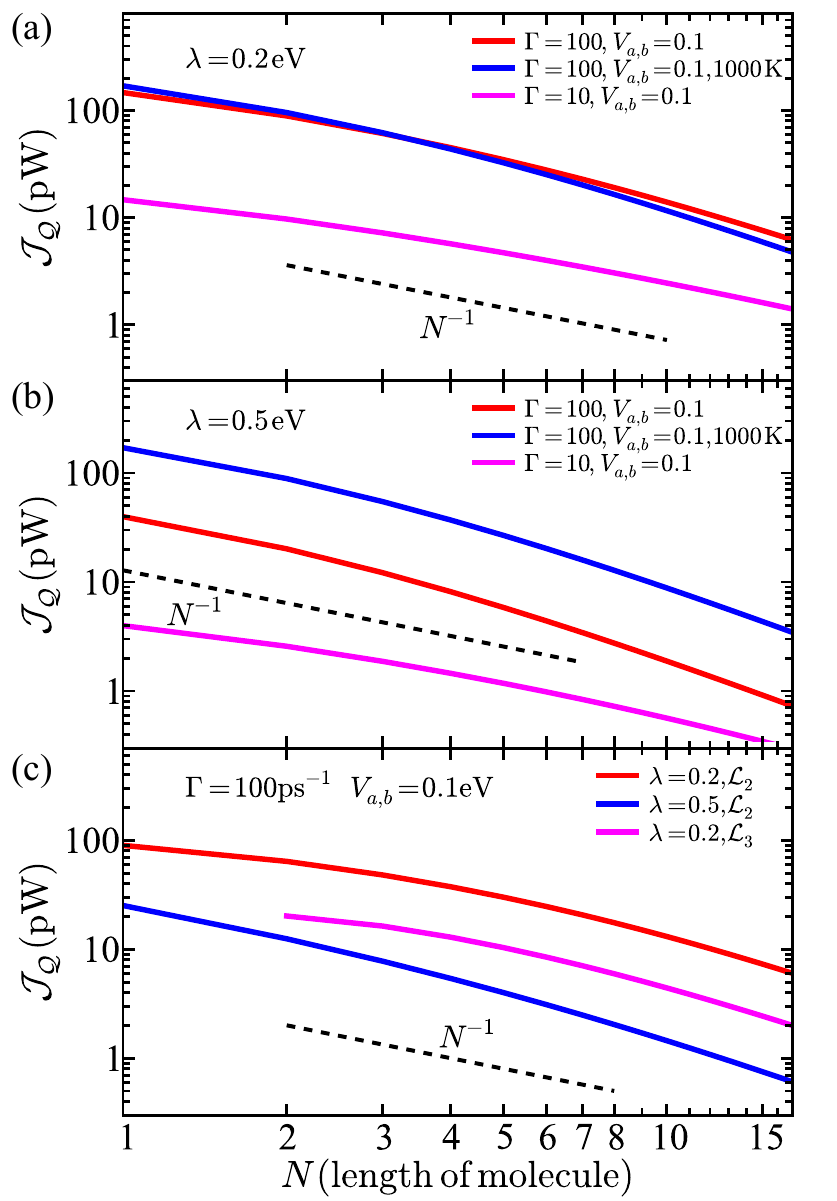}
\caption{\label{fig:HC}
Log-log plot showing heat current $\mathcal{J}_\mathcal{Q}$ as a function of molecular length $N$.
The parameters for each curve are:
$\Gamma = 100\,\text{ps}^{-1}$, $V_{a,b} = 0.1\,\text{eV}$, $T = 300\,\text{K}$ (red),
$\Gamma = 100\,\text{ps}^{-1}$, $V_{a,b} = 0.1\,\text{eV}$, $T = 1000\,\text{K}$ (blue),
$\Gamma = 10\,\text{ps}^{-1}$,  $V_{a,b} = 0.1\,\text{eV}$,  $T = 300\,\text{K}$ (magenta) with 
$\lambda = 0.2\,\text{eV}$ in (a) and $\lambda = 0.5\,\text{eV}$ in (b).
The energy landscape for all curves in (a) and (b) is $\mathcal{L}_1$. 
The bottom panel (c) shows results for energy landscapes $\mathcal{L}_2$ and $\mathcal{L}_3$, both of which are defined in the caption of Fig.~\ref{fig:ER2}.
Other parameters in all panels are $\mu = 0$ and $\Delta T = 1\,\text{K}$.
The dashed curves in each panel show the slope of $N^{-1}$ (Fourier's law) behavior.
}
\end{figure}


Figure~\ref{fig:HC} shows a log-log plot of the ETI heat current as a function of $N$ for various parameter values. 
The dashed lines in each panel show representative $N^{-1}$ scaling that would arise from Fourier's law behavior. 
In Fig.~\ref{fig:HC} (a), results are shown for $\lambda = 0.2\,\text{eV}$ with various combinations of $\Gamma$, $V_{a,b}$, and $T$.
The principal observation is that, in general, the ETI heat current does not follow Fourier's law.
Another important observation is that, in all cases, and even for long molecules, the thermal conductance magnitude is greater than a picowatt. 
In Fig.~\ref{fig:HC}(b), results are shown for $\lambda = 0.5\,\text{eV}$ using the same combinations of $\Gamma$, $V_{a,b}$, and $T$ from Fig.~\ref{fig:HC} (a).
The scaling does scale as $\alpha = -1$ in the regime $N\leq2$ for the red and blue curves, but quickly deviates from this trend for $N>2$. 
For this larger reorganization energy, the heat currents can be sub-$\text{pW}$ as $N$ becomes large in the linear response limit (here with $\Delta T = 1\,\text{K}$).
Figure~\ref{fig:HC}(c) illustrates the heat current behavior as a function $N$ for energy landscapes $\mathcal{L}_2$ and $\mathcal{L}_3$ with various reorganization energies. 
Again, Fourier's law is in general not obeyed. 
This is especially pronounced as $N$ becomes large. 
\begin{figure}
\includegraphics[width = 8.5cm,clip]{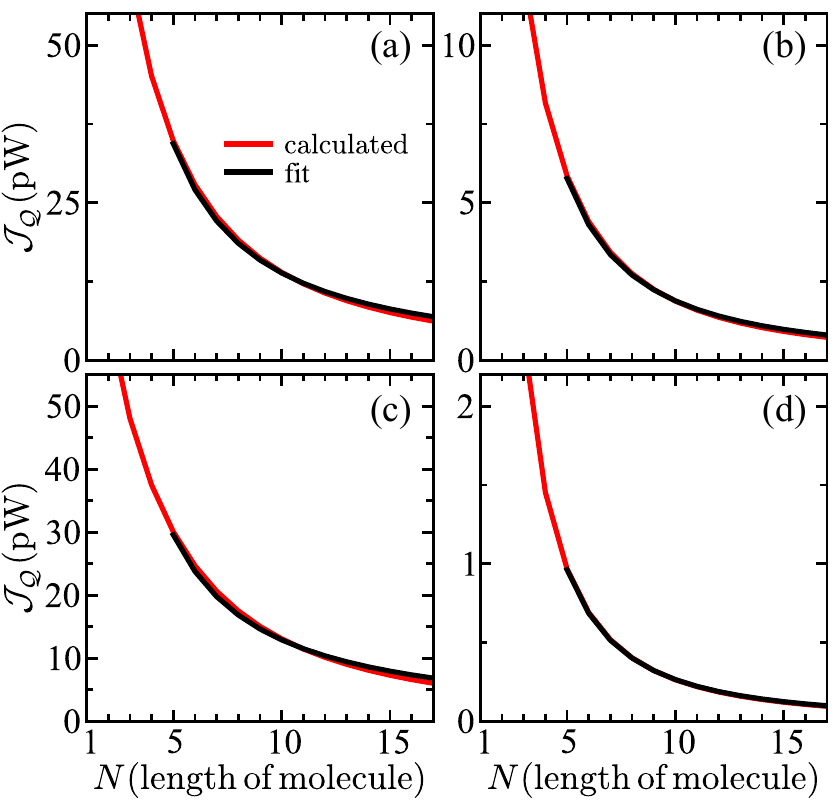}
\caption{\label{fig:HC_fits}
Heat current $\mathcal{J}_\mathcal{Q}$ as a function of molecular length $N$ for 
(a) $\Gamma = 100\,\text{ps}^{-1}$, $V_{a,b} = 0.1\,\text{eV}$, $\lambda= 0.2\,\text{eV}$, 
(b) $\Gamma = 100\,\text{ps}^{-1}$, $V_{a,b} = 0.1\,\text{eV}$, $\lambda= 0.5\,\text{eV}$, 
(c) $\Gamma = 100\,\text{ps}^{-1}$, $V_{a,b} = 0.1\,\text{eV}$, $\lambda= 0.2\,\text{eV}$, $\mathcal{L}_3$
and 
(d) $\Gamma = 10\,\text{ps}^{-1}$, $V_{a,b} = 0.01\,\text{eV}$, $\lambda= 0.2\,\text{eV}$. 
Other parameters in all panels are $T = 300\,\text{K}$, $\Delta T = 1\,\text{K}$, and $\mu = 0$.
The red curve is the calculated value and the black curve is a fit to the functional form $\mathcal{J}_\mathcal{Q} = \mathcal{J}^{(1)}_\mathcal{Q}N^\alpha$.
The energy landscape is $\mathcal{L}_1$ in (a), (b), and (d) and $\mathcal{L}_3$ in (c).
}
\end{figure}

In order to asses validity of Fourier's law to describe ETIHT in long molecules (here defined as molecules with $N \geq 5$ sites), 
we fit the calculated heat current data for $N\geq 5$ to a power law functional form $\mathcal{J}_\mathcal{Q} = \mathcal{J}^{(1)}_\mathcal{Q}N^\alpha$ and extracted values for $\alpha$ from these fits. 
The results for four representative systems, varying $\Gamma$, $V_{a,b}$, and $\lambda$, are shown in Fig.~\ref{fig:HC_fits}. 
The electronic coupling parameter values in Fig.~\ref{fig:HC_fits} (a), (b), and (c) are chosen to represent strong electronic coupling ($\Gamma = 100\,\text{ps}^{-1}$, $V_{a,b} = 0.1\,\text{eV}$) and the values in Fig.~\ref{fig:HC_fits} (d) are chosen to represent weak electronic coupling ($\Gamma = 10\,\text{ps}^{-1}$, $V_{a,b} = 0.01\,\text{eV}$).
In all cases, it can be observed that in the limit of large $N$, fitting the data to a power law provides an acceptable approximation to the decay of the heat current with $N$. 
The scaling factor $\alpha$ can be taken from these fits and compared to the result predicted by Fourier's law. 

\begin{figure}
\includegraphics[width = 8.5cm,clip]{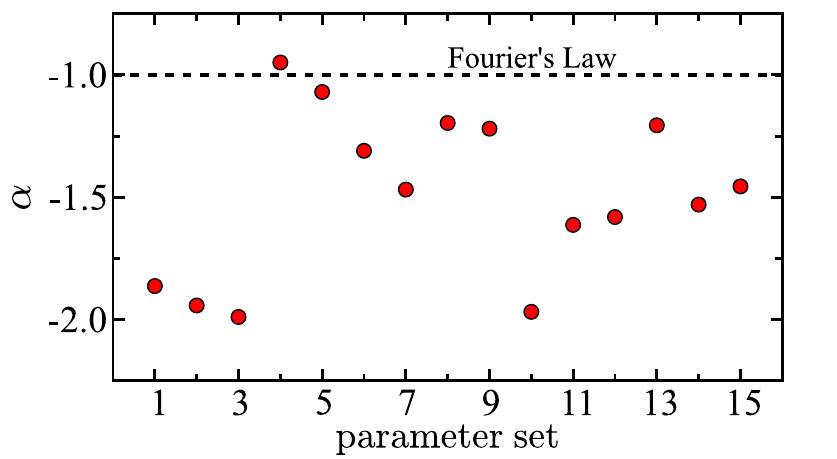}
\caption{\label{fig:alpha}
Scatter plot showing the $\alpha$ values for various parameter sets obtained by fitting the heat currents for molecular lengths of $N \geq 5$ to the functional form $\mathcal{J}_\mathcal{Q} = \mathcal{J}^{(1)}_\mathcal{Q}N^\alpha$. 
The parameter values for each set are listed in Table~\ref{tab:table1} in the Appendix.
The dashed black line denotes behavior that follows Fourier's law ($\alpha = -1$).
}
\end{figure}

Figure~\ref{fig:alpha} shows the calculated values of $\alpha$ for the fifteen parameter sets $\left\{\Gamma, V_{a,b}, \lambda, T, \mathcal{L}\right\}$ listed in Table~\ref{tab:table1} in the Appendix. 
In almost all cases, we observe $\alpha < - 1$. This implies two important points: 
(a) deviations from Fourier's law ($\alpha = -1$) are common and 
(b) because we observe deviations over a broad spectrum of parameter values, this suggests that these deviations arise from the nature of the examined physical mechanism, i.e., ETI heat transport.
We have used parameter values in our calculations that are in agreement with values that have been either observed experimentally or calculated using electronic structure methods and that represent typically experimental conditions. 
We therefore expect that deviations from Fourier's law will be observed in molecules and molecular junctions in which the dominant heat transport mechanism is due to electron hopping.

\section{\label{sec:conc}Conclusions}		

A theoretical formalism has been developed to describe the thermal transport properties generated by electron hopping in molecules and molecular junctions.
This heat transport mechanism has been termed electron-transfer-induced heat transport.
Using this formalism, we have illustrated how the magnitude of electron-transfer-induced heat transport in a molecular junction is affected by varying the electronic properties of the junction's molecular bridge as well as the strength of the coupling between the bridge and its surrounding solvent environment.
We have shown how varying the electronic properties of the molecular bridge alters electron transport and, subsequently, the magnitude of electron-transfer-induced thermal conductance.
The dependence of electron-transfer-induced thermal conductance on the length of the molecular bridge has also been examined, 
and it has been shown that the heat currents generated by electron hopping commonly deviate from Fourier's law behavior.

Comparing the magnitude of the thermal conductance values calculated here with recent experimental measurements in molecular junctions, 
we expect that electron-transfer-induced thermal conductance can make an appreciable and measurable contribution to the overall thermal conductance in properly engineered systems.
Experimental work in that direction is currently underway, with possible applications being to infer values for electron transfer reorganization energies using relationships between thermal and electrical conductance. \cite{Kondratenko2022}

Recently, Sowa \textit{et al.} have derived an expression for molecule-to-metal electron transport rate that interpolates between the Landauer (ballistic) and Marcus (hopping) limits of transport, 
and that also accounts for level broadening. \cite{Sowa2018,Sowa2019} 
The main advantage of this expression is that it can be applied to understand electronic transport properties away from the limiting cases.
Although the validity of this expression outside of the limiting behaviors is unknown, it offers an interesting way to access intermediate transport regimes.
However, it should be noted that in the context of the thermal response properties examined here, this expression is not applicable, principally because using it in heat transfer calculations predicts an unphysically large amount of heat will be transported. 
This is because using this expression gives rise to divergent heat transfer integrals.  
Additionally, the formalism from Refs.\citenum{Sowa2018} and \citenum{Sowa2019} can only describe the case of a single charge transfer site in the molecular bridge, not the case of $N$ sites treated here,
and therefore its applicability is limited in the context of describing transport in long molecules and molecular junctions.


In the theoretical picture developed here, it is assumed that energy relaxation into the solvent environment in response to an electron transfer event occurs instantaneously, and therefore the solvent environment is always in an equilibrium state. 
This physical picture relies on strong assumptions, and recent work that goes beyond this picture to include nonequilibrium effects in the solvent have been developed. \cite{Matyushov2019,Henning2020,Henning2022,Sowa2021}
Extensions of those theories to include the role that soft and complex fluids \cite{Likos06,Ikeda2013,craven14b,Wilding2014,Singh2018}
play in electron transfer reactions may elucidate how structural solvent effects, such as caging in soft matter fluids, affects thermal transport in molecules. 
Work in this direction may lead to enhanced understanding of how tuning the electronic and structural properties of a solvent environment can enhance or suppress thermal transport in junctions for specific technological applications.
In future work, applying theoretical formalisms that treat energy activation and energy relaxation events separately
may give more insight into the physical mechanisms giving rise to electron-transfer-induced heat transport. \cite{craven18a1,craven18a2}

\section{Acknowledgments}
The research of AN is supported by the U.S. National Science Foundation under the Grant No. CHE1953701 and the University of Pennsylvania. 
The research of GTC is supported by Los Alamos National Laboratory through a Laboratory Directed Research and Development (LDRD) grant.

\section{Data Availability Statement}
The data that support the findings of this study are available from the corresponding author upon reasonable request and institutional approval.

\appendix*
\section{\label{sec:param} Parameter Sets}

\begin{table}
\caption{\label{tab:table1}Parameter sets used in Fig.~\ref{fig:alpha}}
\begin{ruledtabular}
\begin{tabular}{ccccccc}
set&$\lambda (\mathrm{eV})$ & $\Gamma (\mathrm{ps^{-1}})$ & $V_{a,b} (\mathrm{eV})$ & $T (\mathrm{K})$ & $\mathcal{L}$ & $\alpha$\\
\cline{1-7}
 1&0.2&10&0.01&300&$\mathcal{L}_1$&-1.86\\
 2&0.2&10&0.01&1000&$\mathcal{L}_1$&-1.94\\
 3&0.5&10&0.01&300&$\mathcal{L}_1$&-1.99\\
 4&0.2&10&0.1&300&$\mathcal{L}_1$&-0.95\\
 5&0.5&10&0.1&300&$\mathcal{L}_1$&-1.07\\
 6&0.2&100&0.1&300&$\mathcal{L}_1$&-1.31\\
 7&0.2&100&0.1&1000&$\mathcal{L}_1$&-1.47\\
 8&0.2&100&0.1&300&$\mathcal{L}_2$&-1.20\\
 9&0.2&100&0.1&300&$\mathcal{L}_3$&-1.22\\
 10&1.0&100&0.1&1000&$\mathcal{L}_1$&-1.97\\
 11&0.5&100&0.1&300&$\mathcal{L}_1$&-1.61\\
 12&0.5&100&0.1&1000&$\mathcal{L}_1$&-1.58\\
 13&0.5&100&0.1&300&$\mathcal{L}_4$&-1.21\\
 14&0.5&100&0.1&1000&$\mathcal{L}_4$&-1.53\\
 15&0.5&100&0.1&1000&$\mathcal{L}_2$&-1.46\\
\end{tabular}
\end{ruledtabular}
\end{table}

\bibliography{c5}

\begin{thebibliography}{78}
\bibitem[{not()}]{note1}
\bibinfo{note}{The molecular junction model considered here is a model we have
  used previously in Refs.~\citenum{craven17a} and \citenum{craven20a}.}
\end{thebibliography}
\end{document}